\documentclass[12pt]{article}

\title{Virial theorem and dynamical evolution  of self-gravitating 
Brownian particles in an unbounded domain:\\
II. Inertial models. }

\usepackage{amsthm,amsmath,amssymb,a4wide,psfig,epsf}
\def\mb#1{\setbox0=\hbox{$#1$}\kern-.025em\copy0\kern-\wd0
\kern-0.05em\copy0\kern-\wd0\kern-.025em\raise.0233em\box0}

\begin{document}

\author{Pierre-Henri Chavanis and Cl\'ement Sire}
\maketitle
\begin{center}
Laboratoire de Physique Th\'eorique (UMR 5152 du CNRS), Universit\'e
Paul Sabatier,\\ 118, route de Narbonne, 31062 Toulouse Cedex 4, France\\
E-mail: {\it chavanis{@}irsamc.ups-tlse.fr \ ; \ clement.sire@irsamc.ups-tlse.fr}

\vspace{0.5cm}
\end{center}

\begin{abstract}

We propose a general kinetic and hydrodynamic description of
self-gravitating Brownian particles in $d$ dimensions. We go beyond
usual approximations by considering inertial effects and finite $N$
effects while previous works use a mean-field approximation valid in a
proper thermodynamic limit ($N\rightarrow +\infty$) and consider an
overdamped regime ($\xi \rightarrow +\infty$). We recover known models
in some particular cases of our general description. We derive the
expression of the Virial theorem for self-gravitating Brownian
particles and study the linear dynamical stability of isolated
clusters of particles and uniform systems by using technics introduced
in astrophysics. We investigate the influence of the equation of
state, of the dimension of space and of the friction coefficient on
the dynamical stability of the system. We obtain the exact expression
of the critical temperature $T_{c}$ for a multi-components
self-gravitating Brownian gas in $d=2$. We also consider the limit of
weak frictions $\xi\rightarrow 0$ and derive the
orbit-averaged-Kramers equation.

\end{abstract}

\section{Introduction}
\label{sec_introduction}

Self-gravitating systems such as globular clusters and galaxies can be
considered as a collection of $N$ stars in gravitational interaction
whose dynamics is described by the Hamilton equations of motion
\cite{bt}. In statistical mechanics, this situation is associated with
the microcanonical ensemble where the energy and the particle number
are fixed \cite{paddy,houches}.  In a recent series of papers
\cite{prs,sc,lang,sic,chs,crrs,sich,sopik}, we have proposed to
consider a system of self-gravitating Brownian particles which are
subject, in addition to the gravitational force, to a friction and a
noise. Their dynamics is described by $N$-coupled stochastic Langevin
equations. In statistical mechanics, this situation is associated with
the canonical ensemble where the temperature and the particle number
are fixed \cite{chav}. In previous papers, we have considered a
mean-field approximation valid in a proper thermodynamic limit with
$N\rightarrow +\infty$ and, in order to simplify the problem, we have
studied a limit of strong friction $\xi\rightarrow +\infty$, or
equivalently a large time regime $t\gg
\xi^{-1}$. In these approximations, the problem is reduced to the
study of the Smoluchowski-Poisson (SP) system. We have also introduced
a generalized class of stochastic processes and kinetic equations in
which the diffusion coefficient (or the friction/mobility) is allowed
to depend on the concentration of particles
\cite{pre,physnext,pa,cban,lemou}. This can model microscopic
constraints (e.g. close packing) acting on the particles when their
local concentration becomes large. The evolution of the system is then
described by the generalized Smoluchowski-Poisson (GSP) system
involving a barotropic equation of state $p(\rho)$ specified by the
stochastic model. This mean-field nonlinear Fokker-Planck (MFNFP)
equation admits a Lyapunov functional, determined by the equation of
state, which can be interpreted as a generalized free energy
\cite{pre}. Thus, this model is associated with a notion of
(effective) ``generalized thermodynamics'' in $\mu$-space.  In the
classical case where the diffusion coefficient is constant, we recover
an isothermal equation of state $p=\rho k_{B}T/m$ associated with the
Boltzmann free energy but more general equations of state can be
considered. Interestingly,  the same type of drift-diffusion
equations are encountered in mathematical biology to describe the
chemotactic aggregration of bacterial populations, in relation with
the Keller-Segel model
\cite{murray,keller,jager}. The analogy between self-gravitating
Brownian particles and bacterial populations is developed in
\cite{crrs}.

Here, we propose to generalize these models so as to take into account
finite $N$ effects and inertial effects (finite friction $\xi$). We
thus propose a general kinetic and hydrodynamic description of
self-gravitating Brownian particles starting directly from a system of
$N$ coupled Langevin equations of motion with long-range attractive
interactions. We shall extend the technics developed in astrophysics
to our problem of self-gravitating Brownian particles.  In particular,
we shall derive the appropriate expression of the Virial theorem for
these systems and study their linear dynamical stability by a method
similar to that developed by Eddington \cite{eddington} and Ledoux
\cite{ledoux} for barotropic stars described by the Euler
equations.  We shall make the link between parabolic and hyperbolic
models by considering an intermediate model taking into account the
inertia of the particles as well as a friction force. The Euler
equations are recovered for $\xi=0$ and the Smoluchowski equation is
obtained in the limit $\xi\rightarrow +\infty$.

The paper is organized as follows.  In Sec. \ref{sec_kin}, we
introduce general kinetic and hydrodynamic models of self-gravitating
Brownian particles starting from coupled Langevin equations. We derive
the $N$-body Fokker-Planck equation (subsection \ref{sec_nbody}), the
mean-field Kramers equation (subsection \ref{sec_kram}) and the
generalized mean-field Kramers equation (subsection \ref{sec_gkram}).
Then, we take the hydrodynamic moments of these equations and derive
the damped Jeans equations (subsection \ref{sec_jeans}) and the damped
barotropic Euler equations (subsection \ref{sec_eul}) by closing the
hierarchy of moments with a local thermodynamical equilibrium
(L.T.E.) hypothesis.  We obtain the mean-field Smoluchowski equation in
a strong friction limit $\xi\rightarrow +\infty$ and, in subsection
\ref{sec_oak}, we derive the orbit-averaged-Kramers equation is a weak
friction limit $\xi\rightarrow 0$. In Sec. \ref{sec_vith}, we
establish the general expression of the Virial theorem for
self-gravitating Brownian particles from the damped Jeans equations
(subsection \ref{sec_vj}) and from the damped Euler equations
(subsection \ref{sec_ve}). We also consider the effect of correlations
due to finite $N$ effects (subsection \ref{sec_corrv}). In
Sec. \ref{sec_dyn}, we study the linear dynamical stability of an
inhomogeneous stationary solution of the damped barotropic
Euler-Poisson system and investigate the effect of the friction
coefficient on the evolution of the perturbation.  In Appendix
\ref{sec_pol} we give a short complement concerning the stability of
polytropic systems and in Appendix \ref{sec_exV} we derive the exact
expression of the Virial theorem starting directly from the stochastic
equations of motion. We show that the Virial theorem takes a very
simple form in dimensions $d=2$ and $d=4$ and analyze the consequences
of this simplification. In Appendix \ref{sec_hom}, we study the linear
dynamical stability of homogeneous stationary solutions of the damped
barotropic Euler equations (for an arbitrary potential of interaction)
and obtain a generalization of the Jeans instability criterion. In the
Conclusion, we discuss the different regimes in the evolution of
Hamiltonian and Brownian systems with long-range interactions \cite{chav}
distinguishing the phase of violent collisionless relaxation, the
collisional evolution due to finite $N$ effects and the
``collisional'' evolution due to {\it imposed} friction and stochastic
forces for Brownian systems.

\section{Kinetic and hydrodynamic models of
self-gravitating Brownian particles}
\label{sec_kin}

The Smoluchowski-Poisson system that has been extensively studied in
previous papers \cite{prs,sc,lang,sic,chs,crrs,sich,sopik} describes a
gas of self-gravitating Brownian particles in a mean-field
approximation (valid for $N\rightarrow +\infty$) and in a strong
friction limit $\xi\rightarrow +\infty$. In this section, we introduce
more general kinetic and hydrodynamic models of self-gravitating
Brownian particles that go beyond these approximations.

\subsection{The $N$-body Fokker-Planck equation}
\label{sec_nbody}

Basically, a system of self-gravitating Brownian particles is
described by the $N$ coupled stochastic equations
\begin{eqnarray}
\label{nb1}
{d{\bf r}_{i}\over dt}={\bf v}_{i},
\end{eqnarray}
\begin{eqnarray}
\label{nb2}
{d{\bf v}_{i}\over dt}=-\xi{\bf v}_{i}-m\nabla_{i}U({\bf
r}_{1},...,{\bf r}_{N})+\sqrt{2D}{\bf R}_{i}(t),
\end{eqnarray}
where $-\xi {\bf v}_{i}$ is a friction force and ${\bf R}_{i}(t)$
is a Gaussian white noise satisfying $\langle {\bf
R}_{i}(t)\rangle={\bf 0}$ and $\langle
{R}_{a,i}(t){R}_{b,j}(t')\rangle=\delta_{ij}\delta_{ab}\delta(t-t')$,
where $a,b=1,...,d$ refer to the coordinates of space and
$i,j=1,...,N$ to the particles. We define the inverse temperature
$\beta=1/k_B T$ through the Einstein relation $\xi=D\beta m$ (see
below). For $\xi=0$ and $D=0$, Eqs.~(\ref{nb1})-(\ref{nb2}) reduce
to the Newton-Hamilton equations of motion describing the ordinary
self-gravitating gas with a Hamiltonian
\begin{equation}
\label{nb3} H=\sum_{i=1}^{N}{1\over 2}m{v_{i}^{2}}+m^{2}U({\bf
r}_{1},...,{\bf r}_{N}),
\end{equation}
where $U({\bf r}_{1},...,{\bf r}_{N})=\sum_{i<j}u({\bf r}_{i}-{\bf
r}_{j})$ and $u({\bf r}_{i}-{\bf r}_{j})=-G/\lbrack (d-2)|{\bf
r}_{i}-{\bf r}_{j}|^{(d-2)}\rbrack$ denotes the gravitational
potential of interaction in $d$ dimensions ($u({\bf r}_{i}-{\bf
r}_{j})=G\ln |{\bf r}_{i}-{\bf r}_{j}|$ for $d=2$). In this paper, we
shall be particularly interested by the gravitational interaction, but
we stress that our formalism remains valid for a more general class of
binary potentials of interaction of the form $u(|{\bf r}_{i}-{\bf
r}_{j}|)$. The evolution of the $N$-body distribution function is
governed by the Fokker-Planck equation
\cite{chav}:
\begin{equation}
\label{nb4} {\partial P_{N}\over\partial t}+\sum_{i=1}^{N}\biggl
({\bf v}_{i}\cdot {\partial P_{N}\over\partial {\bf r}_{i}}+{\bf
F}_{i}\cdot {\partial P_{N}\over\partial {\bf v}_{i}}\biggr
)=\sum_{i=1}^{N} {\partial\over\partial {\bf v}_{i}}\cdot \biggl\lbrack
D{\partial P_{N}\over\partial {\bf v}_{i}}+\xi P_{N}{\bf
v}_{i}\biggr\rbrack.
\end{equation}
In the absence of friction and diffusion ($\xi=D=0$), it reduces
to the Liouville equation. The Liouville equation conserves the
energy  $\langle E\rangle=\int P_{N}H
\prod_i d{\bf r}_{i}d{\bf v}_{i}$ and the Gibbs entropy
$S=-k_{B}\int P_{N}\ln P_{N}d{\bf r}_{1}d{\bf v}_{1}...d{\bf
r}_{N}d{\bf v}_{N}$ (more generally, any functional of $P_{N}$)
defined on $\Gamma$-space. This corresponds to a microcanonical
description.  Alternatively, in the Brownian model, the temperature $T$
is fixed (instead of the energy) and the Fokker-Planck equation
(\ref{nb4}) decreases the Gibbs free energy $F=\langle
E\rangle-TS$ which can be written explicitly,
\begin{equation}
\label{freen} F[P_{N}]=\int P_{N}H
\prod_i d{\bf r}_{i}d{\bf v}_{i}+k_{B}T\int P_{N}\ln P_{N}\prod_i d{\bf r}_{i}d{\bf v}_{i}.
\end{equation}
This corresponds to a
canonical description. One has
\begin{equation}
\label{nb6}
\dot F=-\sum_{i=1}^{N}\int {1\over\xi P_{N}}\biggl (D{\partial
P_{N}\over\partial {\bf v}_{i}}+\xi P_{N}{\bf v}_{i}\biggr )^{2}
d{\bf r}_{1}d{\bf v}_{1}...d{\bf r}_{N}d{\bf v}_{N}\le 0.
\end{equation}
Therefore, the free energy plays the role of a Lyapunov functional for
the $N$-body Fokker-Planck equation.  At equilibrium, $\dot F=0$
implying that the r.h.s. of Eq.~(\ref{nb4}) vanishes. The
l.h.s. (advective term) must also vanish independently. From these
two requirements we find that the
stationary solution of Eq.~(\ref{nb4}) is the canonical distribution
\footnote{In order to properly define a {\it strict} statistical
equilibrium state for self-gravitating systems, one has to introduce a
small-scale regularization otherwise $F[P_{N}]$ has no minimum and
Eq.~(\ref{nb7}) is not normalizable. Thus, in Eq.~(\ref{nb7}), it is
implicitly understood that $U$ is a regularized potential. Note that
{\it physical} statistical equilibrium states unaffected by the
small-scale regularization exist in the form of long-lived metastable
structures that are {\it local} minima of the Boltzmann mean-field
free energy $F_{B}[f]$ defined in Eq. (\ref{fexp}). We refer to
\cite{meta} for a physical discussion of these issues.}
\begin{equation}
\label{nb7} P_{N}({\bf r}_{1},{\bf v}_{1},...,{\bf r}_{N},{\bf
v}_{N})={1\over Z(\beta)}e^{-\beta m
(\sum_{i=1}^{N}{v_{i}^{2}\over 2}+m U({\bf r}_{1},...,{\bf
r}_{N}))},
\end{equation}
provided that the coefficients of diffusion and friction are connected
by the Einstein relation $\xi=D\beta m$. The partition function
$Z(\beta)$ is determined by the normalization condition $\int P_{N}
\prod_i d{\bf r}_{i}d{\bf v}_{i}=1$. The canonical distribution
(\ref{nb7}) minimizes the free energy $F$ at fixed particle
number. Introducing the reduced probability distributions
\begin{equation}
\label{nb8} P_{j}({\bf x}_{1},...,{\bf x}_{j})=\int P_{N}({\bf
x}_{1},...,{\bf x}_{N})\,d{\bf x}_{j+1}...d{\bf x}_{N},
\end{equation}
where ${\bf x}=({\bf r},{\bf v})$, we can readily write down a
hierarchy of equations for $P_{1}$, $P_{2}$ etc. The first
equation of the hierarchy is
\begin{eqnarray}
\label{nb9}\qquad  {\partial P_{1}\over\partial t}+{\bf
v}_{1}\cdot {\partial P_{1}\over\partial {\bf r}_{1}}+(N-1)\int
{\bf F}(2\rightarrow 1)\cdot
 {\partial P_{2}\over\partial {\bf v}_{1}}\,d{\bf r}_{2}d{\bf v}_{2}
={\partial\over\partial {\bf v}_{1}}\cdot \biggl\lbrack D{\partial
P_{1}\over\partial {\bf v}_{1}}+\xi P_{1}{\bf v}_{1}\biggr\rbrack,
\end{eqnarray}
where ${\bf F}(2\rightarrow 1)=-m\partial u_{12}/\partial {\bf r}_{1}=Gm ({\bf r}_{2}-{\bf r}_{1})/|{\bf r}_{2}-{\bf r}_{1}|^{d}$ is the force (by unit of mass) created
by particle $2$ on particle $1$. Note that this equation is exact
(i.e. valid for all $N$) and takes into account the correlations
between the particles encapsulated in the two-body distribution
function $P_2$. For $D=\xi=0$, we recover the first equation of the
BBGKY hierarchy. We shall give the form of the Virial theorem
associated to Eq.~(\ref{nb9}) in Sec. \ref{sec_corrv}. Before that, we
consider the mean-field limit of this equation valid for $N\rightarrow
+\infty$.

\subsection{The mean-field  Kramers equation}
\label{sec_kram}

In a properly defined thermodynamic limit \cite{chav}, we can show
that the cumulant of the two-body correlation function is of order
$1/N$. Therefore, for $N\rightarrow +\infty$, we can implement the
mean-field approximation
\begin{eqnarray}
\label{k1} P_{2}({\bf r}_{1},{\bf v}_{1},{\bf r}_{2},{\bf
v}_{2},t)=P_{1}({\bf r}_{1},{\bf v}_{1},t)P_{1}({\bf r}_{2},{\bf
v}_{2},t)+O(1/N).
\end{eqnarray}
Substituting this result in Eq.~(\ref{nb9}) and introducing the
distribution function $f=NmP_{1}$, we obtain the mean-field Kramers
equation
\begin{eqnarray}
\label{k2} {\partial f\over\partial t}+{\bf v}\cdot {\partial
f\over\partial {\bf r}}-\nabla\Phi \cdot {\partial f\over\partial
{\bf v}} ={\partial\over\partial {\bf v}}\cdot \biggl\lbrack
D{\partial f\over\partial {\bf v}}+\xi f{\bf v}\biggr\rbrack,
\end{eqnarray}
where $\Phi({\bf r},t)=\int u({\bf r}-{\bf r}') \rho({\bf r}',t)d{\bf
r}'$.  This equation is non-local because the potential
$\Phi({\bf r},t)$ is induced by the density $\rho({\bf r},t)=\int f
d{\bf v}$ of the particles composing the whole system (it is not an
external potential).  Thus, for self-gravitating systems,
Eq.~(\ref{k2}) has to be solved in conjunction with the Poisson
equation
\begin{eqnarray}
\Delta\Phi=S_{d}G\rho. \label{pois}
\end{eqnarray} 
In the absence of friction and diffusion ($D=\xi=0$), the mean-field
Kramers equation reduces to the Vlasov equation \cite{bt}. The Vlasov
equation conserves the energy $E={1\over 2}\int f v^{2}d{\bf r}d{\bf
v}+{1\over 2}\int \rho\Phi d{\bf r}$ and the Boltzmann entropy
$S_{B}=-\int (f/m)\ln (f/m) d{\bf r}d{\bf v}$ (more generally all the
functionals of $f$ called the Casimirs) defined on
$\mu$-space. Alternatively, the Kramers-Poisson (KP) system
(\ref{k2})-(\ref{pois}) involves a fixed temperature and decreases the
Boltzmann free energy $F_{B}=E-TS_{B}$ which can be written explicitly
\begin{eqnarray}
F_{B}[f]={1\over 2}\int f v^{2}d{\bf r}d{\bf v}+{1\over 2}\int \rho\Phi d{\bf r}+k_{B}T \int {f\over m}\ln {f\over m} d{\bf r}d{\bf v}.\label{fexp}
\end{eqnarray}
Indeed, one has 
\begin{eqnarray}
\dot F_{B}=-\int {1\over\xi f}\biggl (D{\partial
f\over\partial {\bf v}}+\xi f {\bf v}\biggr )^{2} d{\bf r}d{\bf
v}\le 0. \label{fdotk}
\end{eqnarray}
At equilibrium, $\dot F_{B}=0$ implying that the r.h.s. of Eq.~(\ref{k2})
vanishes. The l.h.s.  (advective term) must also vanish
independently. From these two requirements, and using the Einstein
relation, we find that the stationary solutions of the Kramers-Poisson system (\ref{k2}) correspond to the mean-field Maxwell-Boltzmann distribution 
\begin{eqnarray}
f=A e^{-\beta m ({v^{2}\over 2}+\Phi({\bf r}))} \label{mfmax}
\end{eqnarray}
which has to be solved in conjunction with the Poisson equation
(\ref{pois}). The stable mean-field Maxwell-Boltzmann distribution
minimizes the Boltzmann free energy $F_{B}[f]$ at fixed mass. It is
both thermodynamically stable (in the canonical ensemble) and linearly
dynamically stable with respect to the KP system \cite{pre}. We note
that the equilibrium one-body distribution function (\ref{mfmax}) can
be obtained from the $N$-body canonical distribution (\ref{nb7}) by
constructing an equilibrium BBGKY-like hierarchy and implementing a
mean-field approximation \cite{chav}. On the other hand, the Boltzmann
free energy (\ref{fexp}) can be deduced from the free energy of the
$N$-body system (\ref{freen}) by making the mean-field approximation
$P_{N}(1,...,N)=\prod_{i} P_{1}(i)$ \cite{chav}.

\subsection{The generalized non-local Kramers equation}
\label{sec_gkram}

For sake of generality, we shall consider the case where the diffusion
coefficient explicitly depends on the distribution function. Thus, in
Eq. (\ref{k2}), we set $D(f)=D f C''(f)$ where $C$ is a convex
function, i.e. $C''\ge 0$, and $D$ is a constant. In that case, we obtain
the generalized mean-field Kramers equation \cite{pre}:
\begin{eqnarray}
\label{k3} {\partial f\over\partial t}+{\bf v}\cdot {\partial
f\over\partial {\bf r}}-\nabla\Phi \cdot {\partial f\over\partial {\bf
v}} ={\partial\over\partial {\bf v}}\cdot \biggl\lbrack
DfC''(f){\partial f\over\partial {\bf v}}+\xi f{\bf
v}\biggr\rbrack.
\end{eqnarray}
This equation can be obtained in the mean-field limit of a generalized
$N$-body Fokker-Planck equation associated with a generalized class of
stochastic processes of the form (\ref{nb1})-(\ref{nb2}) where the
diffusion coefficient depends on $f({\bf r}_{i},{\bf v}_{i},t)$
\cite{chav}. For $C=f\ln f$, we recover the
classical Kramers equation (\ref{k2}).  However, Eq.~(\ref{k3}) can
describe more general situations such as quantum particles with
exclusion or inclusion principles (fermions, bosons, quons), lattice
models, non-ideal effects etc...  These generalized Fokker-Planck
equations are associated with an effective thermodynamical formalism
(ETF) in $\mu$-space \cite{pre}. In particular, the generalized
Kramers-Poisson (GKP) system decreases the free energy
\begin{eqnarray}
\label{k4} F[f]\equiv E-TS=\int f{v^{2}\over 2}\,d{\bf r}d{\bf v}+{1\over
2}\int \rho\Phi \,d{\bf r}+T\int C(f)\,d{\bf r}d{\bf v},
\end{eqnarray}
where the last term can be interpreted as a generalized entropy
$S=-\int C(f)d{\bf r}d{\bf v}$ and we have defined the effective
temperature $T=1/\beta$ through the relation $\xi=D/T$ (effective
Einstein relation). One has
\begin{eqnarray}
\label{k5} \dot F=-\int {1\over\xi f}\biggl (DfC''(f){\partial
f\over\partial {\bf v}}+\xi f {\bf v}\biggr )^{2} d{\bf r}d{\bf
v}\le 0.
\end{eqnarray}
At equilibrium, we find that the stationary solutions of the
generalized Kramers equation (\ref{k3}) are determined by the
integro-differential equation
\begin{eqnarray}
\label{k6} C'(f)=-\beta\biggl ({v^{2}\over 2}+\Phi\biggr )-\alpha,
\end{eqnarray}
where $\Phi({\bf r},t)=\int u({\bf r}-{\bf r}') f({\bf r}',{\bf
v}',t)\,d{\bf r}'d{\bf v}'$.  Since $C$ is convex, this relation can
be inversed to give $f=F(\beta\epsilon+\alpha)$ where
$F(x)=(C')^{-1}(-x)$ and $\epsilon={v^{2}\over 2}+\Phi({\bf r})$.  We
note that the equilibrium distribution determined by Eq. (\ref{k6}) is
a function $f=f(\epsilon)$ of the individual energy
$\epsilon$ alone which is monotonically decreasing
(for $\beta>0$).  This equilibrium distribution function extremizes
the free energy (\ref{k4}) at fixed mass. Furthermore, only a (local)
minimum of free energy is linearly dynamically stable with respect to
the GKP system \cite{pre}.

\subsection{The damped Jeans equations}
\label{sec_jeans}

We shall now determine the hierarchy of moment equations
associated with the generalized Kramers-Poisson system. Defining
the density and the local velocity by
\begin{eqnarray}
\label{j1} \rho=\int f\,d{\bf v}, \qquad \rho{\bf u}=\int f{\bf
v}\,d{\bf v},
\end{eqnarray}
and integrating Eq.~(\ref{k3}) on velocity, we get the continuity equation
\begin{eqnarray}
\label{j2} {\partial\rho\over\partial t}+\nabla\cdot (\rho{\bf u})=0.
\end{eqnarray}
Next, multiplying Eq.~(\ref{k3}) by ${\bf v}$ and integrating on velocity, we obtain
\begin{eqnarray}
\label{j3} {\partial\over\partial t}(\rho
u_{i})+{\partial\over\partial x_{j}}(\rho u_{i}u_{j})= -{\partial
P_{ij}\over\partial x_{j}}-\rho{\partial\Phi\over\partial
x_{i}}-\xi\rho u_{i},
\end{eqnarray}
where we have defined the pressure tensor
\begin{eqnarray}
\label{j4}P_{ij}=\int fw_{i}w_{j}\,d{\bf v},
\end{eqnarray}
where ${\bf w}={\bf v}-{\bf u}$ is the relative velocity.  In the absence
of diffusion and friction ($D=\xi=0$), we recover the Jeans equations
of astrophysics that are derived from the Vlasov equation
\cite{bt}. For self-gravitating Brownian particles, the equivalent of
the Jeans equations (\ref{j2})-(\ref{j3}) include an additional
friction force $-\xi {\bf u}$. Using the continuity equation,
Eq.~(\ref{j3}) can be rewritten
\begin{eqnarray}
\label{j5} \rho\biggl ({\partial u_{i}\over\partial
t}+u_{j}{\partial u_{i}\over\partial x_{j}}\biggr )=-{\partial
P_{ij}\over\partial x_{j}}-\rho{\partial\Phi\over\partial
x_{i}}-\xi\rho u_{i}.
\end{eqnarray}

\subsection{The damped barotropic Euler equations}
\label{sec_eul}

By taking the successive moments of the velocity, we can obtain a
hierarchy of hydrodynamical equations. Each equation of the hierarchy
involves the moment of next order. The ordinary Jeans equations that
are based on the Vlasov equation are difficult to close because the
Vlasov equation admits an infinite number of stationary
solutions. Therefore, a notion of thermodynamical equilibrium is
difficult to justify in the usual point of view (see, however,
\cite{csr} in the context of the theory of violent relaxation). In the
present case, the situation is simpler because the Kramers equation
admits a Lyapunov functional (\ref{k4}) and a unique
stationary distribution defined by Eq.~(\ref{k6}).  If we are sufficiently
close to equilibrium, it makes sense to close the hierarchy of
equations by using a condition of local thermodynamic equilibrium. We
shall thus determine the pressure tensor Eq.~(\ref{j4}) with the
distribution function $f_{L.T.E}$ defined by the relation
\begin{eqnarray}
\label{eul1} C'(f_{L.T.E.})=-\beta\biggl \lbrack {w^{2}\over
2}+\lambda({\bf r},t)\biggr \rbrack.
\end{eqnarray}
This distribution function minimizes the generalized free energy
Eq.~(\ref{k4}) at fixed temperature $T$, local density $\rho({\bf r},t)$
and local velocity ${\bf u}({\bf r},t)$. The function
$\lambda({\bf r},t)$ is the Lagrange multiplier associated with
the density field. It is determined by the requirement
\begin{eqnarray}
\label{eul2}\rho({\bf r},t)=\int f_{L.T.E.} \,d{\bf v}=\rho\lbrack
\lambda({\bf r},t)\rbrack.
\end{eqnarray}
At equilibrium, we recover the distribution function
Eq.~(\ref{k6}) with ${\bf u}({\bf r})={\bf 0}$ and $\lambda({\bf
r})=\Phi({\bf r})+\alpha/\beta$. Using the condition
Eq.~(\ref{eul1}) of local thermodynamic equilibrium (LTE), the
pressure tensor Eq.~(\ref{j4}) can be written
$P_{ij}=p\delta_{ij}$ with
\begin{eqnarray}
\label{eul3} p({\bf r},t)={1\over d}\int f_{L.T.E.} w^{2}\,d{\bf
w}=p\lbrack \lambda({\bf r},t)\rbrack.
\end{eqnarray}
The pressure is a function $p=p(\rho)$ of the density which is
entirely specified by the function $C(f)$, by eliminating $\lambda$
from the relations Eq.~(\ref{eul2}) and Eq.~(\ref{eul3}).  We note
furthermore that, using Eq. (\ref{eul1}) and integrations by parts,
the previous equations (\ref{eul2}) and (\ref{eul3}) easily
lead to $\nabla p={1\over d}\int {\partial f\over\partial {\bf
r}}w^{2}d{\bf w}={1\over d}\nabla\lambda
\int {\partial f\over\partial {\bf w}}{\bf w}d{\bf
w}=-\nabla\lambda\int f d{\bf w}= -\rho\nabla
\lambda$, hence $p'(\rho)=-\rho
\lambda'(\rho)$.  In the case of Brownian particles described by the
ordinary Kramers equation (\ref{k2}) with $C(f)=f\ln f$, the equation
of state determined by Eqs. (\ref{eul2}) and (\ref{eul3}) is the
isothermal one $p={k_{B}T\over m}\rho$. More generally, we obtain the damped
Euler equations for a barotropic gas \cite{pre}:
\begin{eqnarray}
\label{eul4} {\partial\rho\over\partial t}+\nabla\cdot (\rho{\bf u})=0,
\end{eqnarray}
\begin{eqnarray}
\label{eul5} {\partial {\bf u}\over\partial t}+({\bf u}\cdot
\nabla){\bf u}= -{1\over\rho}\nabla p-\nabla\Phi-\xi {\bf u}.
\end{eqnarray}
These equations decrease the free energy
\begin{eqnarray}
\label{eul6} F[\rho,{\bf u}]=\int \rho\int^{\rho}{p(\rho')\over
\rho'^{2}} \,d \rho'd{\bf r}+{1\over 2}\int\rho\Phi d{\bf r}+\int
\rho {{\bf u}^{2}\over 2} \,d{\bf r},
\end{eqnarray}
which can be deduced from the free energy (\ref{k4}) by using the
local thermodynamic equilibrium condition (\ref{eul1}) to express
$F[f]$ as as a functional of $\rho$ and ${\bf u}$, using $F[\rho,{\bf
u}]=F[f_{L.T.E.}]$ (see \cite{lemou} for details). A direct
calculation yields
\begin{eqnarray}
\label{eul7} \dot F=-\xi\int \rho {\bf u}^{2}\,d{\bf r}\le 0.
\end{eqnarray}
At equilibrium, $\dot F=0$ implying ${\bf u}={\bf 0}$. Then,
Eq.~(\ref{eul5}) yields the condition of hydrostatic balance 
\begin{eqnarray}
\label{eulhydro} \nabla p+\rho\nabla\Phi={\bf 0},
\end{eqnarray}
which also results from Eq. (\ref{k6}). Indeed, for $f=f(\epsilon)$ with $\epsilon=v^{2}/2+\Phi({\bf r})$, one has $\rho=\int f(\epsilon)d{\bf v}$, $p={1\over d}\int f(\epsilon)v^{2}d{\bf v}$ so that
\begin{eqnarray}
\label{hproof} \nabla p={1\over d}\int f'(\epsilon)\nabla\Phi v^{2}d{\bf v}={1\over d}\nabla\Phi \int {\partial f\over\partial {\bf v}}\cdot {\bf v}d{\bf v}=-\nabla\Phi\int f d{\bf v}=-\rho\nabla\Phi.
\end{eqnarray}

The damped barotropic Euler equations (\ref{eul4})-(\ref{eul5}) are
interesting as they connect hyperbolic models to parabolic
models. Indeed, for $\xi=0$ we recover the standard barotropic Euler
equations describing the dynamics of gaseous stars
\cite{bt,aa1,aa2,grand} or the formation of large-scale structures in
Cosmology \cite{peeble}. Alternatively, in the strong friction limit
$\xi\rightarrow +\infty$, we can neglect the inertial term in
Eq.~(\ref{eul5}) and we obtain
\begin{eqnarray}
\label{eul8} {\bf u}=-{1\over\xi\rho}(\nabla
p+\rho\nabla\Phi)+O(\xi^{-2}).
\end{eqnarray}
Substituting this drift term in the continuity equation (\ref{eul4}), we get
the generalized Smoluchowski equation \cite{pre}:
\begin{eqnarray}
{\partial\rho\over\partial t}=\nabla\cdot \biggl \lbrack
{1\over\xi}(\nabla p+\rho\nabla\Phi)\biggr\rbrack. \label{eul9}
\end{eqnarray}
This equation decreases the free energy 
\begin{eqnarray}
\label{eul6b} F[\rho]=\int \rho\int^{\rho}{p(\rho')\over
\rho'^{2}} \,d \rho'd{\bf r}+{1\over 2}\int\rho\Phi d{\bf r},
\end{eqnarray}
which is obtained from Eq. (\ref{eul6}) by neglecting the last term of
order $O(\xi^{-2})$. A direct calculation yields
\begin{eqnarray}
\label{eul7b} \dot F=-\int {1\over\xi \rho}(\nabla p+ \rho \nabla\Phi)^{2}\,d{\bf r}\le 0.
\end{eqnarray}

It should be recalled that the damped Euler equations
(\ref{eul4})-(\ref{eul5}) remain heuristic because their derivation is
based on a Local Thermodynamic Equilibrium (L.T.E.) assumption
(\ref{eul1}) which is not rigorously justified. However, using a
Chapman-Enskog expansion, it is shown in \cite{lemou} that the
generalized Smoluchowski equation (\ref{eul9}) is {\it exact} in the
limit $\xi\rightarrow +\infty$ (or, equivalently, for times $t\gg
\xi^{-1}$).  The generalized Smoluchowski equation can also be
obtained from the moments equations of the generalized Kramers
equation by closing the hierarchy, using $\xi\rightarrow +\infty$ (see
Sec. 9 of \cite{cban}). 

\subsection{The orbit-averaged-Kramers equation}
\label{sec_oak}

We shall consider here the opposite limit of low friction
$\xi\rightarrow 0$.  In the case where the term in the r.h.s of
Eq. (\ref{k2}) is small, we can obtain a simplified equation for the
evolution of the distribution function by averaging the kinetic
equation over the orbits of the particles. In the case of Hamiltonian
self-gravitating systems described by the Landau equation, this leads
to the orbit-averaged-Landau equation introduced by H\'enon
\cite{henon}. We shall here derive the orbit-averaged-Kramers equation
for self-gravitating Brownian systems.

Let us first rewrite the mean-field Kramers equation (\ref{k2}) in the
form
\begin{eqnarray}
\label{oak1} {\partial f\over\partial t}+{\bf v}\cdot {\partial
f\over\partial {\bf r}}-\nabla\Phi \cdot {\partial f\over\partial
{\bf v}} =Q(f).
\end{eqnarray}
We consider the case where $\xi\rightarrow 0$ for fixed $\beta$ so
that $\xi\simeq D\simeq 0$ and the operator $Q(f)$ can be considered
as small. If we take $Q(f)=0$, we obtain the Vlasov equation. We shall
assume that the system has reached a stable stationary distribution of
the Vlasov equation of the form $f=f(\epsilon)$ which depends only on
the energy $\epsilon={v^{2}\over 2}+\Phi({\bf r},t)$ of the
particles. This is a particular case of the Jeans theorem for
spherical systems. Such steady solution can arise from a process of
violent collisionless relaxation
\cite{lb,chav}. Since $Q(f)\neq 0$, the distribution function 
will change due to the terms of friction and diffusion that are
present in the stochastic equation Eq. (\ref{nb2}) \footnote{In the
case of Hamiltonian systems where $\xi=D=0$, the distribution function
changes due to finite $N$ effects representing close encounters
between stars. These encounters are usually modelled by the Landau
operator
\cite{chav} which is of order $1/N\ll 1$. On the other hand, for self-gravitating Brownian particles the evolution is modeled by 
the Kramers operator (obtained for $N\rightarrow +\infty$) which is of
the order $\xi t_{D}$ (where $t_{D}$ is the dynamical time). The
comparison between the evolution of Hamiltonian and Brownian systems
with long-range interactions is further discussed in the Conclusion of
this paper.}. However, if $\xi\rightarrow 0$, this change will be slow
so that the latter forces cause only a small variation on the
energy. We shall therefore consider that $f({\bf r},{\bf v},t)\simeq
f(\epsilon,t)$ remains a function of the energy alone that slowly
evolves due to imposed stochastic forces. Noting that
\begin{eqnarray}
\label{oak2} {\partial \over\partial t}f(\epsilon({\bf r},{\bf v},t),t)= {\partial f\over\partial t}+{\partial \Phi\over\partial t}{\partial f\over\partial \epsilon},
\end{eqnarray}
we can rewrite the kinetic equation (\ref{oak1}) in the form
\begin{eqnarray}
\label{oak3} {\partial f\over\partial t}+{\partial \Phi\over\partial t}{\partial f\over\partial \epsilon}=Q(f).
\end{eqnarray}
Since $f$ depends only on the energy, the system is spherically symmetric. Then,  the phase space hypersurface with energy smaller than $\epsilon$ is
\begin{eqnarray}
\label{oak4} q(\epsilon,t)\equiv 16\pi^{2}\int_{v^{2}/2+\Phi\le \epsilon}r^{2}dr v^{2}dv={16\pi^{2}\over 3}\int_{0}^{r_{max}(\epsilon,t)}\lbrack 2(\epsilon-\Phi)\rbrack^{3/2}r^{2}dr,
\end{eqnarray}
where $r_{max}(\epsilon,t)$ is the largest radial extent of an orbit with energy $\epsilon$ at time $t$. It is determined by the equation $\epsilon=\Phi(r_{max},t)$ corresponding to $v=0$. The previous relation can be written more compactly as
\begin{eqnarray}
\label{oak5} q(\epsilon,t)={16\pi^{2}\over 3}\int_{0}^{r_{max}} v^{3} r^{2}dr,
\end{eqnarray}
where $v=\sqrt{2(\epsilon-\Phi(r,t))}$. The phase space hypersurface $g(\epsilon,t)d\epsilon$ with energy between $\epsilon$ and $\epsilon+d\epsilon$ is given by
\begin{eqnarray}
\label{oak6} g(\epsilon,t)\equiv {\partial q\over\partial \epsilon}={16\pi^{2}}\int_{0}^{r_{max}(\epsilon,t)}\lbrack 2(\epsilon-\Phi)\rbrack^{1/2}r^{2}dr={16\pi^{2}}\int_{0}^{r_{max}} v r^{2}dr.
\end{eqnarray}
Now, the density of particles in the hypersurface between $\epsilon$
and $\epsilon+d\epsilon$ is uniform since the distribution function
depends only on the energy. In fact, the system evolves on a short
timescale $\sim t_{D}$ by purely inertial effects (corresponding to
the advective terms in the l.h.s of Eq. (\ref{oak1})) so as to
establish this quasi-stationary regime where $f\simeq
f(\epsilon,t)$. We shall therefore average the kinetic equation
(\ref{oak3}) on each hypersurface of iso-energy using
\begin{eqnarray}
\label{oak7} \langle X\rangle (\epsilon,t)={\int_{0}^{r_{max}} Xv r^{2}dr\over \int_{0}^{r_{max}} v r^{2}dr}
\end{eqnarray}
for any function $X(r,v,t)$. Thus, the orbit-averaged kinetic equation can be written
\begin{eqnarray}
\label{oak8}16\pi^{2}\int_{0}^{r_{max}}r^{2}dr \ v\left\lbrack {\partial f\over\partial t}+{\partial\Phi\over\partial t}{\partial f\over\partial\epsilon}-Q(f)\right\rbrack=0.
\end{eqnarray} 
The first term in bracket is 
\begin{eqnarray}
\label{oak9}16\pi^{2}\int_{0}^{r_{max}}r^{2}dr \ v {\partial f\over\partial t}=g(\epsilon,t){\partial f\over\partial t}={\partial q\over\partial \epsilon}{\partial f\over\partial t}.
\end{eqnarray}  
Since
\begin{eqnarray}
\label{oak10}{\partial q\over\partial t}=-16\pi^{2}\int_{0}^{r_{max}}v{\partial\Phi\over\partial t}r^{2}dr,
\end{eqnarray} 
the second term in bracket can be written  
\begin{eqnarray}
\label{oak11}16\pi^{2}\int_{0}^{r_{max}}r^{2}dr \ v {\partial\Phi\over\partial t}{\partial f\over\partial \epsilon}=-{\partial q\over\partial t}{\partial f\over\partial \epsilon}.
\end{eqnarray}  
Finally, since
\begin{eqnarray}
\label{oak12}
vQ(f)=D{\partial\over\partial\epsilon}\left\lbrack v^{3}\left ({\partial f\over\partial \epsilon}+\beta f\right )\right\rbrack,
\end{eqnarray}  
the last term in bracket is
\begin{eqnarray}
\label{oak13}16\pi^{2}\int_{0}^{r_{max}}r^{2}dr \ v Q(f)=3D{\partial\over\partial\epsilon}\left\lbrack q\left ({\partial f\over\partial \epsilon}+\beta m f\right )\right \rbrack.
\end{eqnarray}  
Regrouping all these results, we obtain the orbit-averaged-Kramers equation
\begin{eqnarray}
\label{oak14}{\partial q\over\partial\epsilon}{\partial f\over\partial t}-{\partial q\over\partial t}{\partial f\over\partial \epsilon}=3D{\partial\over\partial\epsilon}\left\lbrack q\left ({\partial f\over\partial \epsilon}+\beta m f\right )\right \rbrack,
\end{eqnarray}  
\begin{eqnarray}
\label{oak15} q(\epsilon,t)={16\pi^{2}\over 3}\int_{0}^{r_{max}(\epsilon,t)}\lbrack 2(\epsilon-\Phi(r,t))\rbrack^{3/2}r^{2}dr,
\end{eqnarray}
\begin{eqnarray}
\label{oak16} {1\over r^{2}}{\partial\over\partial r}\left (r^{2}{\partial\Phi\over\partial r}\right )=16\pi^{2}G\int_{\Phi(r,t)}^{+\infty}f(\epsilon,t)\lbrack 2(\epsilon-\Phi(r,t))\rbrack^{1/2}r^{2}d\epsilon,
\end{eqnarray}
where the last equation is the Poisson equation. It is easy to verify
that the free energy is monotonically decreasing ($\dot F\le 0$) and
that the stationary solution of Eq. (\ref{oak14}) is the Boltzmann distribution
$f=Ae^{-\beta m\epsilon}$. This equations will be studied in a future communication. Note also that in $d=1$, $q=\oint v dx$, $g=2\oint dx/v$ and in $d=2$, $q=2\pi^{2}\int_{0}^{r_{m}}v^{2}rdr$ and $g=2\pi^{2}r_{m}^{2}$.

\section{The Virial theorem for Brownian particles}
\label{sec_vith}

\subsection{The Virial theorem from the damped Jeans equations}
\label{sec_vj}

We shall give here the form of the Virial theorem appropriate to
the damped Jeans equations (\ref{j2})-(\ref{j3}). The only difference with the
standard Jeans equations is the presence of a friction term. We
shall thus only give the final result and refer to \cite{bt} for
the details of calculation. The damped Virial theorem can be
written
\begin{eqnarray}
\label{damv1} {1\over 2}{d^{2}I_{ij}\over dt^{2}}+{1\over 2} \xi
{dI_{ij}\over dt}=2K_{ij}+W_{ij}-{1\over 2}\oint (P_{ik}
x_{j}+P_{jk} x_{i})\,dS_{k}.
\end{eqnarray}
We have included boundary terms which must be taken into account if
the system is confined within a box. The tensor of inertia $I_{ij}$
and the potential energy tensor $W_{ij}$ are defined in Paper I. 
The kinetic energy tensor is defined by
\begin{eqnarray}
\label{damv2} K_{ij}={1\over 2}\int f v_{i}v_{j}\,d{\bf v}.
\end{eqnarray}
It can be written as
\begin{eqnarray}
\label{damv3} K_{ij}=T_{ij}+{1\over 2}\Pi_{ij},
\end{eqnarray}
where
\begin{eqnarray}
\label{damv4}T_{ij}={1\over 2}\int \rho u_{i}u_{j}\,d{\bf r},
\qquad \Pi_{ij}=\int P_{ij}\,d{\bf r}.
\end{eqnarray}
Note that the tensors $K_{ij}$ and $P_{ij}$ depend on the distribution
function $f({\bf r},{\bf v},t)$, not only on hydrodynamic variables.
The scalar Virial theorem takes the form
\begin{eqnarray}
\label{damv6} {1\over 2}{d^{2}I\over dt^{2}}+ {1\over 2}\xi
{dI\over dt}=2K+W_{ii}-\oint P_{ik}x_{i}\,dS_{k},
\end{eqnarray}
where $I$ is the moment of inertia and
\begin{eqnarray}
\label{damv7} K={1\over 2}\int f v^{2}\,d{\bf r}d{\bf v},
\end{eqnarray}
is the kinetic energy. It can be written
\begin{eqnarray}
\label{damv8} K=T+{1\over 2}\Pi,
\end{eqnarray}
where
\begin{eqnarray}
\label{damv9} T={1\over 2}\int \rho {\bf u}^{2}\, d{\bf r}, \qquad
\Pi=d\int p\,d{\bf r}.
\end{eqnarray}
In the absence of diffusion and friction ($D=\xi=0$), we recover the
usual expression of the Virial theorem issued from the Jeans equations
\cite{bt}. For Brownian particles, the novelty is the presence of a
damping term $\xi\dot I$. 

As pointed out in Paper I, the moment of inertia depends on the origin
$O$ of the system of coordinates. Let ${\bf R}(t)=(1/M)\int \rho
{\bf r} d{\bf r}$ denote the position of the center of mass with
respect to the origin $O$. Using the equation of continuity
(\ref{eul4}), we find that $Md{\bf R}/dt={\bf P}$ where ${\bf P}=\int
\rho {\bf u}d{\bf r}$ is the total impulse. Using the Jeans equation
(\ref{j3}), we find that $d{\bf P}/dt=-\xi {\bf P}$. In our case,
there exists an absolute referential $({\cal R})$. Indeed, in writing
Eq. (\ref{nb2}) we have implicitly assumed that our Brownian particles
evolve in a fluid that it motionless. Otherwise, the friction force in
Eq. (\ref{nb2}) has to be modified according to $-\xi ({\bf
v}_{i}-{\bf U})$ where ${\bf U}$ is the velocity of the fluid
\cite{lemou}. We must work therefore in this referential $({\cal
R})$. If we denote by ${\bf P}_{0}$ the initial value of the impulse,
we get ${\bf P}(t)={\bf P}_{0}e^{-\xi t}$. If now ${\bf R}_{0}$
denotes the initial position of the center of mass with respect to
$O$, we find that
\begin{eqnarray}
\label{rmov} {\bf R}(t)={\bf R}_{0}+{{\bf P}_{0}\over M\xi}(1-e^{-\xi t}).
\end{eqnarray}
Therefore, at $t\rightarrow +\infty$ the center of mass has been shifted by a 
quantity ${\bf P}_{0}/M\xi$. In the strong friction limit $\xi\rightarrow +\infty$, we find that the center of mass is motionless (Paper I).

\subsection{The Virial theorem from the damped Euler equations}
\label{sec_ve}

 The Virial
theorem associated to the damped barotropic Euler equations
(\ref{eul4})-(\ref{eul5}) can be deduced from Eq.~(\ref{damv1}) by
using the fact that $P_{ij}=p(\rho)\delta_{ij}$. This yields
\begin{eqnarray}
\label{damv11} {1\over 2}{d^{2}I_{ij}\over dt^{2}} +{1\over 2}\xi
{dI_{ij}\over dt}=2T_{ij}+{1\over d}\Pi \delta_{ij}+W_{ij}-{1\over
2}\oint p(\delta_{ik} x_{j}+\delta_{jk} x_{i})\,dS_{k},
\end{eqnarray}
\begin{eqnarray}
\label{damv12} {1\over 2}{d^{2}I\over dt^{2}}+{1\over 2} \xi {dI\over
dt}=2T+\Pi+W_{ii}-\oint p {\bf r}\cdot d{\bf S},
\end{eqnarray}
where each quantity is now expressed in terms of hydrodynamic
variables.  At equilibrium, if no macroscopic motion is present
($T=0$) and if we can neglect the boundary term, we get
\begin{eqnarray}
\label{damv13} W_{ij}=-{1\over d}\Pi\, \delta_{ij}.
\end{eqnarray}
In the absence of diffusion and friction ($D=\xi=0$), we recover the
usual form of Virial theorem issued from the barotropic Euler
equations \cite{bt}.  Alternatively, in the strong friction limit
$\xi\rightarrow +\infty$, we can neglect the term $\ddot I$ in front
of $\xi\dot I$. Furthermore, since the velocity field scales as ${\bf
u}=O(\xi^{-1})$, the kinetic energy tensor $T_{ij}=O(\xi^{-2})$ can
also be neglected. Therefore, the Virial theorem associated with the generalized Smoluchowski-Poisson system
(\ref{eul9})-(\ref{pois}) can be written
\begin{eqnarray}
\label{damv11b} {1\over 2}\xi
{dI_{ij}\over dt}=2 K_{ij}+W_{ij}-{1\over
2}\oint p(\delta_{ik} x_{j}+\delta_{jk} x_{i})\,dS_{k},
\end{eqnarray}
\begin{eqnarray}
\label{damv12b} {1\over 2} \xi {dI\over
dt}=2 K+W_{ii}-\oint p {\bf r}\cdot d{\bf S},
\end{eqnarray}
where
\begin{eqnarray}
\label{kdef} K_{ij}={1\over d}K\delta_{ij}, \qquad K={d\over 2}\int p d{\bf r},
\end{eqnarray}
is the expression of the kinetic energy  to leading order in the limit $\xi\rightarrow +\infty$ where ${\bf u}({\bf r},t)$ can be neglected. We thus recover the results of Paper I starting directly from the GSP system.

\subsection{The effect of correlations}
\label{sec_corrv}

If we account for the effect of correlations (due to finite values of
$N$) between the particles and use the exact kinetic equation (\ref{nb9}),
we obtain the exact damped Jeans equations
\begin{eqnarray}
\label{corrv1} {\partial\rho\over\partial t}+\nabla\cdot (\rho{\bf u})=0,
\end{eqnarray}
\begin{eqnarray}
\label{corrv2} {\partial\over\partial t}(\rho u_{i})+
{\partial\over\partial x_{j}}(\rho u_{i}u_{j})=-{\partial
P_{ij}\over\partial x_{j}}+Gm^{2}\int {x_{i}'-x_{i}\over |{\bf
r}'-{\bf r}|^{d}}\rho_{2}({\bf r},{\bf r}',t)\,d{\bf r}'-\xi\rho
u_{i},
\end{eqnarray}
where we have introduced the spatial correlation function
\begin{eqnarray}
\label{corrv3} \rho_{2}({\bf r}_{1},{\bf r}_{2},t)=N(N-1) \int
P_{2}({\bf r}_{1},{\bf v}_{1},{\bf r}_{2},{\bf v}_{2},t)\,d{\bf
v}_{1}d{\bf v}_{2}.
\end{eqnarray}
In the mean-field approximation $\rho_{2}(1,2)=\rho(1)\rho(2)$, we
recover the damped Jeans equations (\ref{j2})-(\ref{j3}). The Virial theorem is
now given by
\begin{eqnarray}
\label{corrv4} {1\over 2}{d^{2}I_{ij}\over dt^{2}} +{1\over 2}\xi
{dI_{ij}\over dt}=2K_{ij}+W^{corr}_{ij}-{1\over 2}\oint (P_{ik}
x_{j}+P_{jk} x_{i})\,dS_{k},
\end{eqnarray}
where
\begin{equation}
W_{ij}^{corr}=-{Gm^{2}\over 2}\int \rho_{2}({\bf r},{\bf r}')
{({x}_{i}-{x}'_{i})(x_{j}-x_{j}')\over |{\bf r}-{\bf
r}'|^{d}}\,d{\bf r}d{\bf r}', \label{corrv5}
\end{equation}
is a generalization of the potential energy tensor accounting
for correlations between particles. In the strong friction limit $\xi\rightarrow +\infty$,  the Virial theorem reduces to
\begin{eqnarray}
\label{corrv4str} {1\over 2}\xi
{dI_{ij}\over dt}=\delta_{ij}\int p d{\bf r}+W^{corr}_{ij}-{1\over 2}\oint 
p (\delta_{ik}
x_{j}+\delta_{jk} x_{i})\,dS_{k}.
\end{eqnarray}
If we consider the case of Brownian particles with an isothermal
equation of state $p=\rho k_{B}T/m$ and if we focus on a space
with $d=2$ dimensions where
\begin{equation}
W_{ii}^{corr}=-{Gm^{2}\over 2}\int \rho_{2}({\bf r},{\bf r}')
\,d{\bf r}d{\bf r}'=-N(N-1){Gm^{2}\over 2},
 \label{corrv6}
\end{equation}
the scalar Virial theorem takes the form
\begin{eqnarray}
\label{d2a} 
{1\over 2} \xi {dI\over dt}=2Nk_{B}(T-T_{c})-2PV,
\end{eqnarray}  
with a critical temperature
\begin{eqnarray}
\label{yt2}
k_{B}T_{c}={Gm^{2}(N-1)\over 4}.
\end{eqnarray}
These results are valid for an arbitrary number of
particles.  For $N\gg 1$, using $N-1\simeq N$, we recover the results
of Paper I.

\section{Dynamical stability of self-gravitating Brownian particles}
\label{sec_dyn}

We shall now investigate the linear dynamical stability of a
stationary solution of the damped barotropic Euler-Poisson system
(\ref{eul4})-(\ref{eul5}) satisfying the condition of hydrostatic
balance (\ref{eulhydro}).  We shall determine in particular the
equation of pulsations satisfied by a small perturbation around this
equilibrium state. We shall investigate the influence
of the friction parameter $\xi$ on the pulsation period and make the
connection between the standard Euler-Poisson system (hyperbolic)
obtained for $\xi=0$ and the generalized Smoluchowski-Poisson system
(parabolic) obtained for $\xi\rightarrow +\infty$.  The linearized
damped Euler-Poisson equations are
\begin{eqnarray}
\label{pusl1} {\partial\delta \rho\over\partial t} +\nabla\cdot
(\rho\delta {\bf u})=0,
\end{eqnarray}
\begin{eqnarray}
\label{pusl2} \rho {\partial \delta {\bf u}\over\partial t}=
-\nabla(
p'(\rho)\delta\rho)-\rho\nabla\delta\Phi-\delta\rho\nabla\Phi-\xi\rho
\delta{\bf u},
\end{eqnarray}
\begin{eqnarray}
\label{pusl3} \Delta\delta\Phi=S_{d}G\delta\rho.
\end{eqnarray}
Considering spherically symmetric systems and writing the evolution of
the perturbation as $\delta\rho\sim e^{\lambda t}$, we get
\begin{eqnarray}
\label{pusl4} \lambda\delta\rho+{1\over r^{d-1}}{d\over dr}
(r^{d-1}\rho\delta u)=0,
\end{eqnarray}
\begin{eqnarray}
\label{pusl5} \lambda\rho\delta u=-{d\over dr}(p'(\rho)\delta\rho)-
\rho {d\delta\Phi\over dr}-\delta\rho{d\Phi\over dr}-\xi\rho
\delta{u},
\end{eqnarray}
\begin{eqnarray}
\label{pusl6} {1\over r^{d-1}}{d\over dr}\biggl
(r^{d-1}{d\delta\Phi\over dr}\biggr )=S_{d}G\delta\rho.
\end{eqnarray}
As in Paper I, we introduce the function $q(r)$ defined by
\begin{eqnarray}
\label{nov1} \delta\rho={1\over S_{d}r^{d-1}}{dq\over dr}.
\end{eqnarray}
The continuity equation then yields
\begin{eqnarray}
\label{nov2}\delta u=-{\lambda q\over S_{d}\rho r^{d-1}}.
\end{eqnarray}
After some elementary transformations similar to those of Paper I,
Eq.~(\ref{pusl5}) can be put in the form
\begin{equation}
{d\over dr}\biggl ({p'(\rho)\over S_{d} \rho r^{d-1}}{dq\over dr}\biggr
)+{Gq\over r^{d-1}}={\lambda(\lambda+\xi)\over S_{d} \rho  r^{d-1}} q.
\label{pusl7}
\end{equation}
The case of barotropic stars described by the Euler-Poisson system
corresponds to $\xi=0$ \cite{bt,aa1,aa2,grand}. The case of
self-gravitating Brownian particles described by the generalized
Smoluchowski-Poisson system is recovered for $\xi\gg \lambda$
(see \cite{pre} and Paper I). We can therefore use the results of Paper
I by making the substitution $\xi\lambda\rightarrow
\lambda(\lambda+\xi)$. Therefore, an approximate
analytical expression for the eigenvalue $\lambda$ is given by
\begin{equation}
\lambda(\lambda+\xi)=(d\overline{\gamma}+2-2d)(d-2){W\over I}, \qquad (d\neq 2)
\label{pusl8}
\end{equation}
\begin{equation}
\lambda(\lambda+\xi)=-(\overline{\gamma}-1){GM^{2}\over I}, \qquad (d=2).
\label{pusl9}
\end{equation}
The friction coefficient $\xi$ affects the
evolution of the instability but it does not change the
instability threshold (determined by the sign of the l.h.s. of Eqs.
(\ref{pusl8})-(\ref{pusl9})). The unstable case corresponds to
$\lambda(\lambda+\xi)=\sigma^{2}>0$. The two eigenvalues are
\begin{equation}
\lambda_{\pm}={-\xi\pm \sqrt{\xi^{2}+4\sigma^{2}}\over 2}.
\label{pusl10}
\end{equation}
Since $\lambda_{+}>0$, we see that the perturbation grows
exponentially rapidly as $e^{\lambda_{+}t}$. The stable case corresponds to
$\lambda(\lambda+\xi)=-\sigma^{2}<0$. The two eigenvalues are
\begin{equation}
\lambda_{\pm}={-\xi\pm \sqrt{\xi^{2}-4\sigma^{2}}\over 2}.
\label{pusl11}
\end{equation}
If $\xi^{2}-4\sigma^{2}\ge 0$, then $\lambda_{\pm}<0$ and the
perturbation decreases exponentially rapidly without oscillating. This
is the case in particular for Brownian particles described by the
Smoluchowski equation ($\xi\rightarrow +\infty$) for which
$\lambda=-\sigma^{2}/\xi$ (Paper I). Alternatively, if
$\xi^{2}-4\sigma^{2}\le 0$, then $\lambda_{\pm}=(-\xi\pm
i\sqrt{4\sigma^{2}-\xi^{2}})/2$ and we have slowly damped oscillations
with a pulsation $\omega={1\over 2}\sqrt{4\sigma^{2}-\xi^{2}}$ and a
damping rate $\xi/2$. This is the case in particular for barotropic
stars ($\xi=0$) which oscillate with pulsation $\omega=\sigma$ without
attenuation. The separation between these two regimes (pure
damping vs damped oscillations) is obtained for $\xi=2\sigma$ at which
$\omega=0$. This suggests to introducing the dimensionless parameter
\begin{equation}
F\equiv {\xi^{2}\over \lambda(\lambda+\xi)},
\label{pusl12}
\end{equation}
measuring the efficiency of the friction force. The critical values
are $F=0$ and $F=-4$. If $F<-4$ the system is stable and a
perturbation is damped out exponentially rapidly without oscillating.  If
$-4<F<0$ the system is stable and a perturbation exhibits damped
oscillations. The pulsation vanishes for $F=-4$ and the damping rate
vanishes for $F=0$.  For $F>0$, the system is unstable. Using
Eqs.~(\ref{pusl8})-(\ref{pusl9}), the parameter defined in
Eq.~(\ref{pusl12}) is explicitly given by
\begin{equation}
F={1\over (3\overline{\gamma}-4)}{\xi^{2} I\over W},\qquad (d=3)
\label{pusl13}
\end{equation}
\begin{equation}
F={-1\over \overline{\gamma}-1}{\xi^{2} I\over GM^{2}},\qquad (d=2)
\label{pusl14}
\end{equation}
\begin{equation}
F=-{1\over \overline{\gamma}}{\xi^{2} I\over W},\qquad (d=1).
\label{pusl15}
\end{equation}
Dimensionally, this parameter scales as $|F|\sim \xi^{2}R^{d}/GM$. It
can also be written $|F|\sim (\xi t_{D})^{2}$ where $t_{D}\sim
1/\sqrt{\rho G}$ is the dynamical time \cite{bt}. The dynamical
stability of a homogeneous system (for a general form of potential of
interaction) is treated in Appendix \ref{sec_hom}.

\section{Conclusion}
\label{sec_conclusion}

In this paper, we have introduced general models of self-gravitating
Brownian particles (stochastic $N$ body, kinetic, hydrodynamic,...) 
that relax the simplifying assumptions that are usually considered:
mean-field approximation for $N\rightarrow +\infty$ (thermodynamic
limit) and overdamped approximation for $\xi\rightarrow +\infty$
(strong friction limit). These general models show the connection
between previously considered models and offer a unifying framework to
study these systems. We have focused here on the case of
self-gravitating systems but most of our results also apply to the
problem of chemotaxis in biology. This will be specifically considered
in another paper where we discuss inertial models of bacterial
populations.

It should be emphasized that the Brownian model
(\ref{nb1})-(\ref{nb2}) contains the standard Hamiltonian model of
stellar dynamics \cite{bt} as a special case since the Langevin
equations reduce to the Hamilton equations for $\xi=D=0$. We expect
therefore to have different regimes depending on the value of the
parameters. To characterize these regimes properly, it is useful to
introduce different timescales: (i) the {\it dynamical time}
$t_{D}\sim 1/\sqrt{\rho G}$ (Kepler time) is the typical period of an
orbit or a typical free-fall time \cite{bt} (ii) the {\it collisional
relaxation time} $t_{R}\sim (N/\ln N) t_{D}$ (Chandrasekhar time) is
the typical time it takes a stellar system (Hamiltonian) to relax to
the Boltzmann distribution $e^{-\beta m\epsilon}$ due to close
encounters. This relaxation is due to finite $N$ effects
\cite{chandras} (iii) the {\it friction time} $t_{B}\sim \xi^{-1}$ (Kramers time) is the typical time it takes a Brownian system to thermalize, i.e. to
have its velocity distribution close to the Maxwellian $e^{-\beta m
v^{2}/2}$ \cite{risken}. This thermalization is due to the combined
effect of imposed friction and diffusion in the Langevin model
(\ref{nb1})-(\ref{nb2}). It is due to a thermal bath (of
non-gravitational origin) {\it not} to collisions (finite $N$
effects). We can now distinguish different cases:

(1) The case $t_{D}\ll t_{R} \ll t_{B}$ ($\xi\rightarrow 0$)
corresponds to Hamiltonian systems. For $t\ll t_{R}$, the system is
described by the Vlasov-Poisson system. There is first a phase of {\it
violent collisionless relaxation} on a timescale $\sim t_{D}$ leading
to a quasi-stationary state (QSS) in mechanical equilibrium. This is a
stable stationary solution of the Vlasov equation (on the
coarse-grained scale) that is usually not described by the Boltzmann
distribution. On longer timescale $t_{R}\sim (N/\ln N)t_{D}$ the
encounters between stars (due to finite $N$ effects) have the tendency
to drive the system towards a statistical equilibrium state described
by the Boltzmann distribution. In reality, this process is hampered by
the escape of stars and the gravothermal catastrophe. The collisional
evolution of the system is described by the Landau-Poisson system
which is the $1/N$ correction to the Vlasov limit (it singles out the
Boltzmann distribution among all stationary solutions of the Vlasov
equation)\cite{chav}. In fact, due to the time scale separation
between the phase of violent relaxation (inertial effects) and the
phase of collisional relaxation (finite $N$ effects), we can consider
for intermediate times that the distribution function is a
quasi-stationary solution of the Vlasov equation of the form
$f=f(\epsilon,t)$ (for spherical systems) that slowly evolves under
the action of close encounters according to the orbit-averaged-Landau
equation (traditionally called orbit-averaged-Fokker-Planck
equation). This implies that the lifetime of the QSS is {\it long} as
it increases as a power of $N$. It {\it slowly} evolves under the
effect of encounters which act as a perturbation of order $1/N$ with
$N\gg 1$.  Therefore, the system first reaches a state of mechanical
equilibrium (through violent relaxation) then a state of thermal
equilibrium (through stellar encounters). These different phases of
the dynamical evolution of Hamiltonian stellar systems have been
studied by astrophysicists for a long time \cite{bt}.

(2) The case $t_{B}\ll t_{D}\ll t_{R}$ ($\xi\rightarrow +\infty$)
corresponds to the overdamped limit of the Brownian model. The
velocities first relax towards the Maxwellian distribution on a
timescale $t_{B}\sim \xi^{-1}$ (due to the thermal bath) and the
density relaxes towards a state of mechanical equilibrium on a longer
timescale (Smoluchowski diffusive time). Therefore, the system first
reaches a state of thermal equilibrium (because of the terms of
friction and noise in the Langevin equations) then a state of
mechanical equilibrium (through inertial effects). This overdamped
regime, described by the Smoluchowski-Poisson system, has been studied
in our series of papers
\cite{prs,sc,lang,sic,chs,crrs,sich,sopik}.

(3) Finally, there is an interesting case $t_{D}\ll t_{B}\ll t_{R}$
that has not yet been studied. In that case, there is first a phase of
violent relaxation on a timescale $\sim t_{D}$ leading to a
quasi-stationary state (QSS) in mechanical equilibrium like in case
(1). This phase is followed by a thermalization leading to the
Boltzmann distribution on a timescale $t_{B}\sim \xi^{-1}$ due to the
thermal bath, i.e. the combined effect of imposed friction and
diffusion in the Langevin model (\ref{nb1})-(\ref{nb2}), {\it not} to
``collisions'' (finite $N$ effects) as in case (1). The first phase is
described by the Vlasov-Poisson system and the second phase by the
Kramers-Poisson system. For $\xi\rightarrow 0$ (but $\xi\gg (\ln
N/N)t_{D}^{-1}$) there is a time scale separation between the phase of
violent relaxation and the phase of Brownian relaxation. Similarly to
case (1), we can consider for intermediate times that the distribution
function is a quasi-stationary solution of the Vlasov equation of the
form $f=f(\epsilon,t)$ (for spherical systems) that slowly evolves
under the action of imposed friction and diffusion (thermal bath, not
collisions) according to the orbit-averaged-Kramers equation derived
in Sec. \ref{sec_oak}. Since the Brownian timescale $t_{B}\sim
\xi^{-1}$ is independent on $N$, this implies that the lifetime of the
QSS in this regime is independent on $N$. Furthermore, it is {\it
shorter} than in case (1) if $\xi\gg (\ln N/N)t_{D}^{-1}$.  Therefore,
the system first reaches a state of mechanical equilibrium (through
violent relaxation) then a state of thermal equilibrium (through the
effect of imposed fluctuation and dissipation, i.e. the thermal
bath). This is the opposite situation to case (2). The study of the
orbit-averaged-Kramers equation will be considered in a future
work. Note that if $t_{B}$ and $t_{R}$ are comparable, one must take
into account simultaneously the effect of the thermal bath (friction
and random force) and the effect of collisions (finite $N$
effects). This is another interesting case. Finally, we stress that
these different regimes should be observed for other potentials of
interaction $u(|{\bf r}-{\bf r}'|)$ than the gravitational one
(e.g. for the HMF and BMF models \cite{cvb}). Kinetic theories of
Hamiltonian and Brownian particles with long-range interactions are
discussed in \cite{chav} at a general level.

\vskip1cm

{\bf Acknowledgements} We are grateful to the referees for useful remarks that helped to improve the presentation of the paper.

\appendix

\section{Eigenvalue equation for polytropes}
\label{sec_pol}

In the case of a polytropic equation of state, we can put the
eigenvalue equation (\ref{pusl7}) in a dimensionless form \cite{aa2}:
\begin{equation}
{d\over d\xi}\biggl ({\theta^{1-n}\over \xi^{d-1}}{dq\over d\xi}\biggr )
+{nq\over \xi^{d-1}}={n\Omega^{2}\over \theta^{n}\xi^{d-1}}q
\end{equation}
where  $\theta(\xi)$ in the Emden function \cite{chandra} and
\begin{equation}
\Omega^{2}={\lambda (\lambda+\xi)\over S_{d}G\rho_{0}},
\end{equation}
is the dimensionless eigenvalue. This relation shows that the
dimensional eigenvalue scales as $\lambda(\lambda+\xi)\propto
G\rho_{0}$ where $\rho_{0}$ is the central density. Now, for a
polytrope, the central density is connected to the mean density
$\overline{\rho}$ by a proportionality constant depending only on
the polytropic index $n$ \cite{chandra}. Therefore, we also have
$\lambda(\lambda+\xi)\propto G\overline{\rho}$.

\section{The exact Virial theorem in $d$ dimensions}
\label{sec_exV}

We consider $N$ Brownian particles with mass $m_{\alpha}$ in gravitational
interaction. Their equations of motion are
\begin{equation}
\ddot x_{i}^{\alpha}=\sum_{\beta\neq \alpha}
{Gm_{\beta}(x_{i}^{\beta}-x_{i}^{\alpha})\over |{\bf
r}_{\beta}-{\bf r}_{\alpha}|^{d}}-\xi \dot
x_{i}^{\alpha}+\sqrt{2D_{\alpha}}R_{i}^{\alpha}(t). \label{exV0}
\end{equation}
Here, the greak letters refer to the particles and the latin letters
to the coordinates of space.  For simplicity, we have assumed that the
friction coefficient $\xi$ is the same for all the particles but this
assumption can be relaxed easily (see below). The diffusion
coefficient is given by the Einstein formula $D_{\alpha}=\xi
k_{B}T/m_{\alpha}$. The multi-species Smoluchowski-Poisson system has
been studied in
\cite{sopik}. In this Appendix, we establish the exact Virial theorem
associated with the stochastic equations (\ref{exV0}). The moment of
inertia tensor is defined by
\begin{equation}
I_{ij}=\sum_{\alpha}m_{\alpha}x_{i}^{\alpha}x_{j}^{\alpha}.
\label{exV1}
\end{equation}
We introduce the kinetic energy tensor
\begin{equation}
K_{ij}={1\over 2}\sum_{\alpha}m_{\alpha}{\dot x}_{i}^{\alpha}{\dot x}_{j}^{\alpha}
\label{exV2}
\end{equation}
and the potential energy tensor
\begin{eqnarray}
W_{ij}=G\sum_{\alpha\neq\beta} m_{\alpha}m_{\beta}{x_{i}^{\alpha}
(x_{j}^{\beta}-x_{j}^{\alpha})\over |{\bf r}_{\beta}
-{\bf r}_{\alpha}|^{d}}\nonumber\\
=-{1\over 2}G \sum_{\alpha\neq\beta}
m_{\alpha}m_{\beta}{(x_{i}^{\alpha}-x_{i}^{\beta})
(x_{j}^{\alpha}-x_{j}^{\beta})\over
|{\bf r}_{\beta}-{\bf r}_{\alpha}|^{d}}, \label{exV3}
\end{eqnarray}
where the second equality results from simple algebraic manipulations
obtained by exchanging the dummy variables $\alpha$ and $\beta$.
Taking the second derivative of Eq.~(\ref{exV1}), using the equation
of motion (\ref{exV0}), and averaging over the noise and on
statistical realizations, we get
\begin{eqnarray}
{1\over 2} \langle {\ddot I}_{ij}\rangle+{1\over 2}\xi \langle {\dot
I}_{ij}\rangle =2\langle K_{ij}\rangle+\langle W_{ij}\rangle -{1\over 2}\oint
(P_{ik}x_{j}+P_{jk}x_{i})\,dS_{k}, \label{exV4}
\end{eqnarray}
where we have included the Virial of the pressure force
$F_{i}=P_{ik}\Delta S_{k}$ on the boundary of a box: $(\ddot
I_{ij})_{box}=\sum_{box}\sum_{\alpha}(F_{i}^{\alpha}x_{j}^{\alpha}
+F_{j}^{\alpha}x_{i}^{\alpha})=\sum_{box}\sum_{\alpha}(P_{ik}^{\alpha}
x_{j}^{\alpha}+P_{jk}^{\alpha}x_{i}^{\alpha})\,\Delta S_{k}$
\footnote{Note that if the particles have a different friction
parameter $\xi_{\alpha}$, the term $\xi\dot I$ is replaced by
$\sum_{\alpha}\xi_\alpha m_{\alpha}(\dot
x_{i}^{\alpha}x_{j}^{\alpha}+x_{i}^{\alpha}\dot x_{j}^{\alpha})$ which
can also be written $\sum_s\xi_{s}\dot I_{ij,s}$ where $(s)$
denotes the different species of particles and $I_{ij,s}$ is the moment of inertia due to particles of species $s$.}.  The scalar Virial
theorem is obtained by contracting the indices
\begin{eqnarray}
{1\over 2} \langle {\ddot I}\rangle +{1\over 2}\xi \langle {\dot I}\rangle =2\langle K\rangle +\langle W_{ii}\rangle -\oint P_{ik}x_{i}dS_{k},
\label{exV5}
\end{eqnarray}
where
\begin{eqnarray}
I=\sum_{\alpha}m_{\alpha}r_{\alpha}^{2},\qquad K={1\over 2}
\sum_{\alpha}m_{\alpha}v_{\alpha}^{2},
\label{exV6}
\end{eqnarray}
are the moment of inertia and the kinetic energy. On the other hand,
\begin{eqnarray}
W_{ii}=-{1\over 2}G\sum_{\alpha\neq\beta} {m_{\alpha}m_{\beta}\over
|{\bf r}_{\beta}-{\bf r}_{\alpha}|^{d-2}}.
\label{exV7}
\end{eqnarray}
For $d\neq 2$, we find that
\begin{eqnarray}
W_{ii}=(d-2)W
\label{exV8}
\end{eqnarray}
where $W$ is the potential energy
\begin{eqnarray}
W=-{G\over 2(d-2)}\sum_{\alpha\neq\beta}{m_{\alpha}m_{\beta}\over
|{\bf r}_{\beta}-{\bf r}_{\alpha}|^{d-2}}. \label{exV9}
\end{eqnarray}
In that case, the scalar Virial theorem reads
\begin{eqnarray}
{1\over 2}\langle {\ddot I}\rangle +{1\over 2}\xi \langle {\dot I}\rangle =2\langle K\rangle +(d-2)\langle W\rangle -\oint P_{ik}x_{i}dS_{k}.
\label{exV5bis}
\end{eqnarray}
For Hamiltonian systems ($D=\xi=0$), the total energy  $E=K+W$ is
conserved. Hence, the Virial theorem can be written in an
unbounded domain ($P=0$):
\begin{eqnarray}
{1\over 2} {\ddot I}=2K+(d-2)W=2E+(d-4)W.
\label{exV5tris}
\end{eqnarray}
This is the extension of the usual Virial theorem in $d$ dimensions
(this equation is exact without averages). We note that the dimension
$d=4$ is critical as also noticed in \cite{fermionsd} using different
arguments. In that case, ${\ddot I}=4E$ which yields after integration
$I=2Et^{2}+C_{1}t+C_{2}$. For $E>0$, $I\rightarrow +\infty$ indicating
that the system evaporates. For $E<0$, $I$ goes to zero in a finite
time, indicating that the system forms a Dirac peak in a finite
time. More generally, for $d\ge 4$ since $(d-4)W\le 0$ we have $I\le
2Et^{2}+C_{1}t+C_{2}$ so that the system forms a Dirac peak in a
finite time if $E<0$ (this remains true for a box-confined system with
a pressure force at the wall). Therefore, self-gravitating systems
with $E<0$ are not stable in a space of dimension $d\ge 4$. The study
in Ref. \cite{fermionsd} indicates that this observation remains true
even if quantum effects (Pauli exclusion principle) are taken into
account. This is a striking result because quantum mechanics
stabilizes matter against gravitational collapse in $d<4$
\cite{pt}. For $2< d\le 4$, since $(d-4)W\ge 0$ we conclude, according
to (\ref{exV5tris}), that the system evaporates if $E>0$ while an
equilibrium is possible (but not compulsory) if $E<0$. Finally, for
$d<2$, since $W>0$, the energy is necessary positive ($E>0$). Since
$(d-4)W< 0$ an equilibrium state is possible.

For $d=2$, we have the simple result
\begin{eqnarray}
W_{ii}=-{1\over 2}G\sum_{\alpha\neq\beta}m_{\alpha}m_{\beta}.
\label{exV10}
\end{eqnarray}
For equal mass particles,
\begin{eqnarray}
W_{ii}=-{1\over 2}G N(N-1)m^{2}
\label{exV11}
\end{eqnarray}
which reduces to 
\begin{eqnarray}
W_{ii}=-{G M^{2}\over 2},
\label{exV11b}
\end{eqnarray}
 for $N\gg 1$. We also note that
\begin{eqnarray}
\sum_{\alpha\neq\beta}m_{\alpha}m_{\beta}=M^{2}-\sum_{\alpha}m_{\alpha}^{2}.
\label{exV12}
\end{eqnarray}
The first term is of order $N^{2} m^{2}$ and the second of order $N
m^{2}$ (where $m$ is a typical mass). Therefore, in the mean-field
limit $N\rightarrow +\infty$, we recover Eq.~(\ref{exV11b}) even for a
multi-components system, cf
\cite{sopik}.

At equilibrium, the Virial theorem (\ref{exV5}) reduces to 
\begin{eqnarray}
2\langle K\rangle +\langle W_{ii}\rangle =\oint P_{ik}x_{i}dS_{k}.
\label{exV5new1}
\end{eqnarray}
If the system is at statistical equilibrium, then $\langle K\rangle
={d\over 2}Nk_{B}T$ and $P_{ij}=p\delta_{ij}$ with
$p=\sum_{s}\rho_{s}k_{B}T/m_{s}$ \cite{sopik}. Then, introducing the
notation (26) of Paper I, we get
\begin{eqnarray}
dNk_{B}T +\langle W_{ii}\rangle =dPV.
\label{exV5new2}
\end{eqnarray}
For an ideal gas without self-gravity $(W_{ii}=0$), we recover the perfect gas law $PV=Nk_{B}T$. Alternatively, for a self-gravitating gas in two dimensions, using Eq. (\ref{exV11b}), we get the exact equation of state
\begin{equation}
PV=Nk_{B}(T-T_{c}),
\label{exV17n}
\end{equation}
with the {\it exact} critical temperature:
\begin{equation}
k_{B}T_{c}={G \sum_{\alpha\neq\beta}m_{\alpha}m_{\beta}\over 4N}.
\label{exV18}
\end{equation}
For equal mass particles, we get
\begin{equation}
k_{B}T_{c}={Gm^{2}\over 4}(N-1),
\label{exV18b}
\end{equation}
as in Eq.~(\ref{yt2}). In the mean-field limit
\begin{equation}
k_{B}T_{c}={GM^{2}\over 4N}.
\label{exV18bty}
\end{equation}

We now consider the strong friction limit $\xi\rightarrow +\infty$. To
leading order in $1/\xi$, the $N$-body distribution is given by
\begin{equation}
P_{N}({\bf r}_{1},{\bf v}_{1},...,{\bf r}_{N},{\bf v}_{N},t)=e^{-\beta  \sum_{\alpha=1}^{N}m_{\alpha}{v_{\alpha}^{2}\over 2}}\Phi_{N}({\bf r}_{1},...,{\bf r}_{N},t)+O(1/\xi),
\label{fgkdl}
\end{equation}
as can be deduced from the multi-species $N$-body Fokker-Planck
equation \cite{sopik} generalizing Eq. (\ref{nb4}) [this distribution
is obtained by requiring that the Fokker-Planck collision term remains
finite when $D,\xi\rightarrow +\infty$]. From this expression, we find
that $P_{ij}=p\delta_{ij}$ with $p=\sum_{s}\rho_{s}k_{B}T/m_{s}$ and
$\langle K_{ij}\rangle ={1\over 2}Nk_{B}T\delta_{ij}$ even for the
out-of-equilibrium problem. From Eq. (\ref{exV4}), we obtain the
overdamped Virial theorem for a self-gravitating Brownian gas
\begin{equation}
{1\over 2}\xi \langle {\dot I}_{ij}\rangle=\langle W_{ij}\rangle +Nk_{B}T\delta_{ij}-{1\over
2}\oint p(\delta_{ik}x_{j}+\delta_{jk}x_{i})\,dS_{k}.
\label{exV15}
\end{equation}
We can obtain this result from a different manner. In the strong
friction limit $\xi\rightarrow +\infty$, the inertial term in
Eq. (\ref{exV0}) can be neglected so that the stochastic equations of
motion reduce to
\begin{equation}
\dot x_{i}^{\alpha}=\mu_{\alpha}m_{\alpha}\sum_{\beta\neq \alpha}
{Gm_{\beta}(x_{i}^{\beta}-x_{i}^{\alpha})\over |{\bf
r}_{\beta}-{\bf
r}_{\alpha}|^{d}}+\sqrt{2D'_{\alpha}}R_{i}^{\alpha}(t),
\label{exV13}
\end{equation}
where $D'_{\alpha}=k_{B}T\mu_{\alpha}$ and $\mu_{\alpha}=1/\xi m_{\alpha}$. Taking the
derivative of the tensor of inertia (\ref{exV1}) and using
Eq.~(\ref{exV13}), we get
\begin{equation}
{\dot I}_{ij}={2\over\xi} W_{ij}+\sum_{\alpha}
m_{\alpha}\sqrt{2D'_{\alpha}}\lbrack
x_{i}^{\alpha}R_{j}^{\alpha}(t)+x_{j}^{\alpha}R_{i}^{\alpha}(t)\rbrack.
\label{exV14}
\end{equation}
Now, averaging over the noise using $\langle
x_{i}^{\alpha}R_{j}^{\alpha}\rangle=\sqrt{2D'_{\alpha}}\delta_{ij}$,
and on statistical realizations, we recover Eq. (\ref{exV15}). The scalar
Virial theorem reads
\begin{equation}
{1\over 2}\xi \langle {\dot I}\rangle =dNk_{B}T+\langle W_{ii}\rangle -dPV,
\label{exV16}
\end{equation}
where $P$ is defined as in Paper I.  This is similar to Eq.~(32) of
Paper I obtained from the SP system but now $W_{ii}$ is given by
Eq.~(\ref{exV7}). In particular, in $d=2$, using Eq.~(\ref{exV10}) we
obtain
\begin{equation}
{1\over 2}\xi \langle {\dot I}\rangle =2Nk_{B}(T-T_{c})-2PV,
\label{exV17}
\end{equation}
with the {\it exact} critical temperature (\ref{exV18}).

\section{Dynamical stability of homogeneous systems}
\label{sec_hom}

In this Appendix, we study the linear dynamical stability of a
stationary solution of the damped barotropic Euler equations
(\ref{eul4})-(\ref{eul5}) that is infinite and homogeneous, i.e. $\rho_{0}({\bf
r})=\rho$ and ${\bf u}_{0}={\bf 0}$. For sake of generality, we
consider a potential of the form
\begin{equation}
\Phi({\bf r},t)=\int u({\bf r}-{\bf r}')\rho({\bf r}',t)d{\bf r}',
\label{hom1}
\end{equation}
where $u(|{\bf r}-{\bf r}'|)$ is an arbitrary binary potential of
interaction.  We shall thus obtain a generalization of the Jeans
instability criterion. We note that an infinite  homogeneous medium is a
stationary solution of the damped barotropic Euler equations provided that it
satisfies the condition of hydrostatic balance $\nabla
p_{0}+\rho_{0}\nabla\Phi_{0}={\bf 0}$ which reduces to
$\nabla\Phi_{0}={\bf 0}$. With Eq. (\ref{hom1}) this can be written
\begin{equation}
\int {\partial u\over\partial {\bf x}}d{\bf x}={\bf 0} \qquad {\rm or}\qquad \int u(x)d{\bf x}<\infty. 
\label{hom2}
\end{equation}
We shall assume that this condition is fulfilled. We note that for the
gravitational potential, this condition is not fulfilled since
$\nabla\cdot \int \nabla u d{\bf x}=\int \Delta u d{\bf x}=S_{d}G\int
\delta({\bf x})d{\bf x}=S_{d}G\neq 0$. Still, the equations for the
perturbation are well-posed mathematically and, neglecting the
above-mentioned inconsistency at zeroth-order, can be considered as a
first step to investigate the dynamical stability of a gravitational
system. This is the so-called {\it Jeans swindle} \cite{bt}. The
linearized damped barotropic Euler equations can be written
\begin{eqnarray}
\label{hom3} {\partial\delta \rho\over\partial t} +\rho\nabla\cdot
\delta {\bf u}=0,
\end{eqnarray}
\begin{eqnarray}
\label{hom4} \rho {\partial \delta {\bf u}\over\partial t}=
-c_{s}^{2}\Delta\delta \rho-\rho\Delta\delta\Phi-\xi\rho
\delta{\bf u},
\end{eqnarray}
\begin{eqnarray}
\label{hom5} \delta\Phi({\bf r},t)=\int u({\bf r}-{\bf r}')\delta \rho({\bf r}',t)d{\bf r}',
\end{eqnarray}
where we have introduced the velocity of sound $c_{s}^{2}=p'(\rho)$. They can be combined to give
\begin{eqnarray}
\label{hom6}{\partial^{2}\delta\rho\over\partial t^{2}}+\xi{\partial \delta\rho\over\partial t}=c_{s}^{2}\Delta\delta\rho+\rho\Delta\delta\Phi,
\end{eqnarray}
with $\delta\Phi=u * \delta\rho$. Looking for solutions of the form 
\begin{eqnarray}
\label{hom7}\delta\rho({\bf r},t)=\int \delta\hat{\rho}({\bf k},\omega)e^{i({\bf k}\cdot {\bf r}-\omega t)}d{\bf k}d{\omega},
\end{eqnarray}
 we obtain the general dispersion relation
\begin{eqnarray}
\label{hom8}\omega(\omega+i\xi)=c_{s}^{2}k^{2}+(2\pi)^{d}\hat{u}(k)\rho k^{2}.
\end{eqnarray}
Introducing $\lambda=-i\omega$ and $\hat{v}(k)=-(2\pi)^{d}\hat{u}(k)$, this can be rewritten
\begin{eqnarray}
\label{hom8b}\lambda^{2}+\xi\lambda-k^{2}(\rho\hat{v}(k)-c_{s}^{2})=0. 
\end{eqnarray}
The solutions are $\lambda_{\pm}={1\over 2}(-\xi\pm\sqrt{\Delta})$ with
$\Delta(k)=\xi^{2}+4k^{2}(\rho\hat{v}(k)-c_{s}^{2})$. If
$c_{s}^{2}<\rho\hat{v}(k)$, then $\Delta(k)>\xi>0$ and the system is
unstable as $\lambda_{+}={1\over 2}(-\xi+\sqrt{\Delta})>0$. If
$c_{s}^{2}<\rho\hat{v}(k)$, either $\Delta(k)<0$ and
$R_{e}(\lambda)=-\xi/2$ or $0<\Delta(k)<\xi$ implying
$\lambda_{\pm}<0$, so the system is stable. Therefore, the system is stable 
if
\begin{eqnarray}
\label{hom9}c_{s}^{2}>\rho\hat{v}(k),
\end{eqnarray}
and unstable otherwise. For attractive potentials $\hat{v}(k)>0$, this gives rise to the existence of a critical point as discussed in \cite{chav}. Indeed, a necessary condition of instability is that
\begin{eqnarray}
\label{hom10}c_{s}^{2}<(c_{s}^{2})_{crit}=\rho \hat{v}(k)_{max}.
\end{eqnarray}
If this condition is fulfilled the range of unstable wavelengths is determined by 
\begin{eqnarray}
\label{hom11} \hat{v}(k)>{c_{s}^{2}/\rho},
\end{eqnarray}
and their growth rate is $\lambda_{+}(k)$. 
For the Euler equations ($\xi=0$), the dispersion relation reduces to
\begin{eqnarray}
\label{hom13}\omega^{2}=c_{s}^{2}k^{2}+(2\pi)^{d}\hat{u}(k)\rho k^{2}.
\end{eqnarray}
In the unstable case the perturbation grows exponentially and in the
stable case the perturbation presents undamped oscillations. In the
overdamped limit of the model $\xi\rightarrow +\infty$ (Smoluchowski),
the dispersion relation reduces to
\begin{eqnarray}
\label{hom14}i\xi\omega=c_{s}^{2}k^{2}+(2\pi)^{d}\hat{u}(k)\rho k^{2},
\end{eqnarray}
which can be obtained directly from the generalized Smoluchowski
equation (\ref{eul9}).  In the unstable case the perturbation grows
exponentially and in the stable case the perturbation decreases
exponentially. The intermediate case of finite friction can be treated
as in Sec. \ref{sec_dyn} but explicit results demand to specify the
potential of interaction.

For the attractive Yukawa
potential $\hat{v}(k)=S_{d}G/(k^{2}+k_{0}^{2})$ \cite{chav}, we find
that the system is unstable if
$c_{s}^{2}<(c_{s}^{2})_{crit}=S_{d}G\rho/k_{0}^{2}$ for the
wavevectors such that 
\begin{eqnarray}
\label{hom11b} k<k_{max}\equiv \sqrt{{S_{d}G\rho\over c_{s}^{2}}-k_{0}^{2}}.
\end{eqnarray}
The growth rate $\lambda_{+}(k)$ is maximum for $k_{*}^{2}=(S_{d}G\rho k_{0}^{2}/c_{s}^{2})^{1/2}-k_{0}^{2}$ and its value $\lambda_{*}=\lambda_{+}(k_{*})$ is given by
\begin{eqnarray}
\label{hom11c} 2\lambda_{*}=-\xi+\sqrt{\xi^{2}+4S_{d}G\rho \left (1-\sqrt{c_{s}^{2}k_{0}^{2}/ S_{d}G\rho}\right )^{2}}.
\end{eqnarray}
For $c_{s}=0$ (cold systems), we have $k_{max}=k_{*}=+\infty$ and
$\lambda_{*}={1\over 2}(-\xi+\sqrt{\xi^{2}+4S_{d}G\rho})$. The system
is most unstable at small wavelengths. For $k_{0}=0$ (gravitational
potential), we have $(c_{s}^{2})_{crit}=+\infty$, $k_{max}=k_{J}=
\left ({S_{d}G\rho/ c_{s}^{2}}\right )^{1/2}$ (Jeans length),
$k_{*}=0$ and $\lambda_{*}={1\over
2}(-\xi+\sqrt{\xi^{2}+4S_{d}G\rho})$. The system is most unstable at
large wavelengths. For $\xi=0$ (Euler),
$\lambda_{*}=(S_{d}G\rho)^{1/2}(1-\sqrt{c_{s}^{2}k_{0}^{2}/S_{d}G\rho})$
and for $\xi\rightarrow +\infty$ (Smoluchowski),
$\lambda_{*}=(S_{d}G\rho/\xi)(1-\sqrt{c_{s}^{2}k_{0}^{2}/S_{d}G\rho})^{2}$. Other
examples of potentials of interaction are given in \cite{chav}.

\newpage

\end{document}